# On the theory of Biot-patchy-squirt mechanism for wave propagation in partially saturated double-porosity medium

Weitao Sun (孙卫涛)

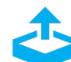

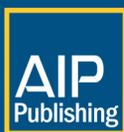

# On the theory of Biot-patchy-squirt mechanism for wave propagation in partially saturated double-porosity medium

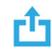

Weitao Sun (孙卫涛)[a])

AFFILIATIONS

School of Aerospace Engineering, Tsinghua University, Beijing 100084, China and Zhou Pei-Yuan Center for Applied Mathematics, Tsinghua University, Beijing 100084, China

[a)]Author to whom correspondence should be addressed: sunwt@tsinghua.edu.cn

ABSTRACT

Reservoir rocks have a coherent heterogeneous porous matrix saturated by multiple fluids. At long wavelength limit, the composite material of solid skeleton is usually regarded as homogeneous media. However, at grain scale or high loading rate, non-uniform fluid flow plays an essential role in wave dispersion and attenuation. Formulating wave propagation in partially saturated and fractured rocks is challenging and is of great interest in geoscience. Recent studies have shown that the mechanisms of wave attenuation caused by viscous dissipation, patchy-saturation, and squirt flow are different. However, the relationship among these mechanisms and the combined effect on wave attenuation are not clear. Here, a Biot-patchy-squirt (BIPS) model is proposed to characterize wave dispersion/attenuation in fractured poroelastic media saturated by immiscible fluids. BIPS model incorporates local fluid-interface flow (LFIF) and squirt flow into global fluid flow simultaneously. Theoretical analysis shows that BIPS is consistent with the Biot theory, squirt flow, and LFIF models, and is reduced to these models under extreme conditions. More interestingly, numerical simulations reveal that the existence of squirt flow partially counterbalances the dissipative effect of LFIF at the patch interface. The attenuation-frequency relationship observed in experiments capturing evidence of squirt flow and patchy-saturation interface flow is reproduced by using the BIPS model. The results show that BIPS model is computationally reliable and is in reasonably good agreement with laboratory data. The findings advance understanding of the physics of wave propagation in natural reservoir rocks and push forward the potential applications of the triple dispersion/attenuation mechanism to wave velocity prediction.

Published under an exclusive license by AIP Publishing. https://doi.org/10.1063/5.0057354

## I. INTRODUCTION

In the last decades, wave dispersion and attenuation in natural reservoir rocks have attracted more and more attention.[1–16] Intrinsic attenuation is believed to result mainly from fluid flow in saturated rocks, which is driven by pressure gradient and obeys Darcy's law. Although the theory of wave propagations in a single porosity and fully saturated porous medium has been well established since Biot,[17,18] energy dissipation mechanisms in tight rock with low porosity/permeability remain poorly understood. Experimental observations have questioned the general validity of single dispersion/attenuation mechanism caused by the dissipation of fluid–solid viscous friction in the single porosity hypothesis.[19,20]

Natural reservoir rocks are usually heterogeneous porous media composed of mineral grains and cements. When the inhomogeneity produces scattering, the fluid–solid interface produces friction, and the squirt flow occurs in fracture/crack, the P-wave will have dispersion and attenuation. The change of pore/fracture structure is an important source of matrix heterogeneity. Barenblatt et al.[21] studied the double-porosity continuum model with fluid exchange between the matrix pore and fracture. Multi-porosity models were proposed based on mass-momentum-energy balance equations by Aifantis,[22–24] Wilson and Aifantis,[25] Khaled et al.,[26] and Elsworth and Bai.[27] Bai et al.[28,29] studied the deformation-dependent flow models. Berryman and Wang[5,30] formulated wave propagations in a double-porosity medium. Chapman incorporated meso-scale anisotropic fracturing into the grain scale fluid flow model.[31] Venegas et al.[32] investigated sound propagation in multiscale porous media with the mass transfer process. They found that the macroscopic mass balance is significantly modified by sorption and inter-scale diffusion. Minale[33,34] studied momentum balance in homogeneous and heterogeneous porous media, providing a way to formulate the stress boundary condition at the interface between a porous medium and a homogeneous fluid.

In addition to solid skeleton heterogeneity, immiscible fluid patches are another important source of attenuation. In oil recovery, time-lapse seismic monitoring, and $CO_2$ injection, the acoustic properties of rock saturated by two immiscible fluids are of great interest. Much effort has been made to set up the quantitative relationships between wave velocities and fluid saturation.[1,6,7,14,35–41] Lattice calculation method has been used for simulation of immiscible fluid displacement in homogeneous and randomly heterogeneous pore networks.[42] Experimental observations have been reported about the velocity-saturation relation in sonic and ultrasonic frequency bands.[43–49] Wave-induced fluid flow due to mesoscopic heterogeneity is believed as one of the main mechanisms underlining the energy dissipation in partially saturated rocks.[6,8,10,40] What is more, fluid patch size and spatial distribution are important factors controlling wave dispersion and attenuation.[10,50–53] The effects of randomly distributed fluid patches have also been studied by Toms *et al.*[11] The coexistence of solid–fluid heterogeneity has been studied by a triple-layer patchy (TLP) model.[14] Experimental results show that injection rate controls the patch formation.[54–56]

Consider a saturated rock with fractures and cracks, the fissured part has greater deformations than the inter-granular pores, which results in a squirt flow from cracks to pores. It is believed that attenuation can be dominated by the squirt flow between pores of different shapes and orientations at sonic and ultrasonic frequencies.[57] Microscopic squirt flow mechanism has been developed to explain the measured attenuation and velocity dispersion.[3,31,58–62] Evidence of squirt flow in sandstone has been reported based on the laboratory observation of seismic attenuation.[63] Comparisons between experimental data and simulation results imply that at ultrasonic frequencies the squirt flow is due to the presence of clay.[64] In bifurcated fractures, flow velocity may be several orders of magnitude greater than the velocity in the matrix.[65] In addition, non-linear flow and reverse flow may form at higher Reynolds numbers in fracture with rough wall.[66] In a highly heterogeneous and fractured porous matrix, the flow and distribution of immiscible fluids remain poorly understood.[67]

On one hand, Biot-squirt (BISQ) model in single-porosity rocks is based on Biot's theory and squirt flow mechanism.[3,61] On the other hand, wave propagation in a partially saturated double-porosity matrix with heterogeneous patches is formulated based on both Biot theory and local fluid-interface flow (LFIF) mechanisms.[5–7,12,14] However, to our knowledge, detailed work on the relationship among global fluid flow (GFF), LFIF, and squirt flow is still missing. Despite the importance from both the theoretical and applied points of view, the laws governing immiscible two-phase flow through heterogeneous double-porosity (pore-fracture) materials are still uncovered.

It is recognized that Biot global flow, squirt flow, and LFIF are distinct from each other. The patchy-saturation effect can be weakened by squirt flow effect in sandstones.[6] Relative importance of global and squirt flow has also been investigated.[68] Recently, the velocity-saturation relations obtained for various injection rates show that patch size may increase with fluid saturation and decrease with injection rate.[56] Since squirt flow is closely related to injection rate, the results imply that patch saturation and squirt flow may have uncovered relation with each other. Unfortunately, how the dispersion/attenuation depends on the interactions of global flow, LFIF, and squirt flow remains poorly understood.

The main goal of this study is to formulate Biot-patchy-squirt (BIPS) flow for wave propagations in partially saturated double-porosity media. Here, the three important energy dissipation sources, say Biot's theory, patchy-saturation, and squirt flow, are believed to be intrinsically interconnected at multiple spatial scales and full range of timescale. Therefore, the aim of this work is to establish a consistent formulation connecting Biot's theory, patchy-saturation, and squirt flow mechanisms in a well-defined way.

The formulation in this work has two main parts. (1) The wave equations of Biot-patchy-squirt mechanism are formulated based Lagrangian equations, mass conservation equations, and constitutive equation of double-porosity media. (2) Consistency among BIPS, Biot, BISQ, and patchy-saturation models is analyzed under extreme conditions. In the numerical example section, P-wave velocity has been predicted for oil-brine saturated sandstone. Compared with experimental data, the BIPS model successfully captures the P-wave velocity dispersion, and the predicted velocities are in reasonably good agreement with laboratory observations.

## II. METHOD
### A. Dynamic equations of Biot-patchy model

Biot-patchy model incorporates solid–fluid heterogeneities by introducing patches embedded in a homogeneous host medium. In a cubic representative element volume (REV), the solid matrix is partitioned into inclusion and host parts by interface $S_{solid}$ [Fig. 1(a)]. The host region is represented as an equivalent concentric shell. The heterogeneous matrix have double-porosity, i.e., the inner spheroid inclusion region with porosity $\phi_{in}$ and the surrounding host region with porosity $\phi_{host}$. The radii of inclusions are much smaller than the wavelength but much larger than the pore size. The interface $S_{solid}$ between inclusions and host medium is open and fluids can flow through it.

Two immiscible fluids saturate the double-porosity skeleton. An interface $S_{fluid}$ exists between the inner fluid pocket and the outside fluid zone [Fig. 1(a)]. The fluid interface $S_{fluid}$ does not necessarily overlap $S_{solid}$. In fact, the solid matrix and pore fluid patch are two independent heterogeneous systems and need to be treated separately. The inner fluid pocket may cover across $S_{solid}$. Nevertheless, the surrounding matrix saturated by the other fluid can also contain both porosities when $S_{fluid}$ is inside $S_{solid}$. Based on such a triple-layer patchy (TLP) model, wave equations have been obtained.[14]

The kinetic energy of the TLP system is

$$T = \frac{1}{2}\sum_{i=1}^{3}\rho_{00}\dot{u}_i^2 + \sum_{i=1}^{3}\sum_{m=1}^{2}\rho_{0m}\dot{u}_i\dot{U}_i^{(m)} + \frac{1}{2}\sum_{i=1}^{3}\sum_{m=1}^{2}\rho_{mm}\dot{U}_i^{(m)2}$$
$$+ \frac{\phi_1\phi_2^2}{6}\dot{\zeta}^2\chi_1, \qquad (1)$$

where $\dot{u}$ is the velocity of solid phase. $U_i^{(m)}$ is the fluid displacements in the $m$th fluid regions and $\chi_1 = a_{20}^2 v_1^{2/3}[\rho_{f_1} + \rho_{f_2}(\phi_{10}/\phi_{20})] \cdot (1 - a_{10}/a_{20})$. Here, $m=1$ denotes the inner fluid pocket and $m=2$ is the surrounding fluid patch. $\zeta = \nabla \cdot \mathbf{Z}$ is the increment of local fluid flow caused by the pressure difference between different fluid patches. The vector $\mathbf{Z}$ represents local fluid flow between immiscible fluid patches. For the heterogeneous matrix, the porosities of inner solid inclusion and surrounding solid background material are $\phi_{10}, \phi_{20}$. The phase volume fraction relation is $\phi_m = v_m\phi_{m0}$ ($m=1, 2$), where $v_1$ is the volume faction of the inclusion and $v_2 = 1 - v_1$ is that of the surrounding background material. $u_i$ is the mesoscopic volume average displacement of solid skeleton. The density coefficients are $\rho_{00} = \rho_0 - \sum_{m=1}^{2} \rho_{f_m}\phi_m(1-1/\phi_{m0})/2$, $\rho_{mm} = \rho_{f_m}\phi_m(1+1/\phi_{m0})/2$, and $\rho_{0m} = \rho_{f_m}\phi_m(1-1/\phi_{m0})/2$, where $\rho_0 = \sum_{m=1}^{2} v_m(1-\phi_{m0})\rho_{s_m}$ is the density of solid skeleton, and $\rho_{f_m}$ are the fluid densities in the $m$th type of fluid patch. $a_{10}$ and $a_{20}$ are the radii of concentric patches.

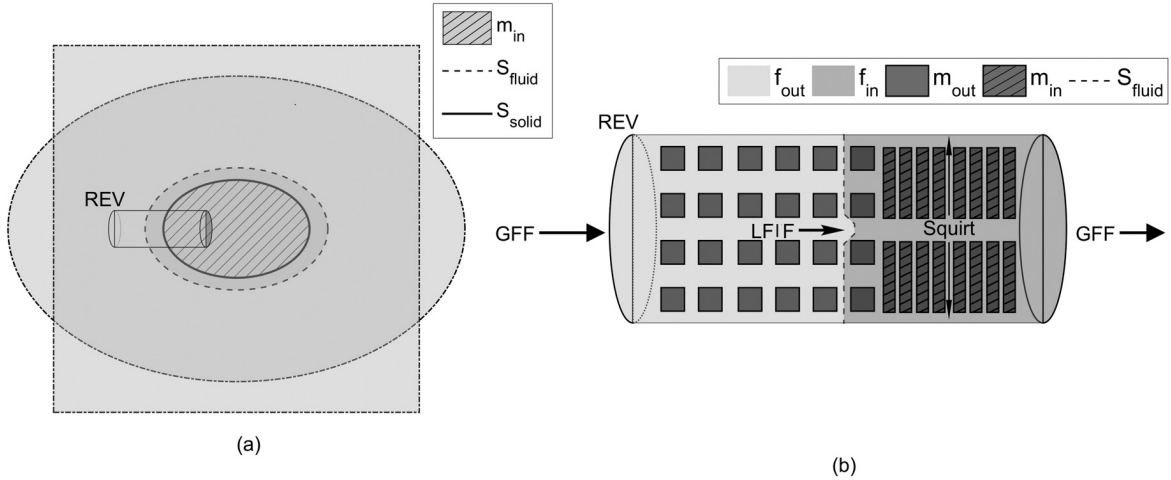

FIG. 1. Schematic diagram of global fluid flow, patchy-saturation, and squirt flow in a heterogeneous porous medium. (a) Biot-patchy model with solid heterogeneities $m_{in}$ (hatched patches) partially saturated by immiscible fluid $f_{out}$ (background) and $f_{in}$ (dark shaded patches). The outside cubic patch volume is represented by an equivalent spheroid. $S_{solid}$ is the solid heterogeneity interface. $S_{fluid}$ is the fluid heterogeneity interface. (b) Biot-patchy-squirt model with a cylindrical representative element volume (REV) containing double porosities and two immiscible fluids. Global fluid flow (GFF) occurs at larger scale level. Local fluid-interface flow (LFIF) occurs at the patch interface. Squirt flow occurs in the solid heterogeneities $m_{in}$ (hatched patches).

The strain energy is

$$W = \frac{1}{2}\big[(A+2N)I_1^2 - 4NI_2 + 2Q_1I_1(\xi_1 - \phi_2\zeta) \\ + R_1(\xi_1 - \phi_2\zeta)^2 + 2Q_2I_1(\xi_2 + \phi_1\zeta) + R_2(\xi_2 + \phi_1\zeta)^2\big], \quad (2)$$

where $I_1$ and $I_2$ are the first and second strain invariants of the solid matrix. $\xi_m$ are the first strain invariants of the average fluid displacement $\mathbf{U}^{(m)}$ ($m=1, 2$). $N$ is the average shear modulus of the solid matrix. Stiffness coefficients $A, Q_m, R_m$ ($m = 1, 2$) are calculated from the porosity $\phi$, the frame bulk modulus $K_b$, and the solid and fluid bulk moduli $K_s, K_f$.[17]

The dissipation function is

$$D = \frac{1}{2}\sum_{m=1}^{2} b_m \sum_{i=1}^{3}\left(\dot{U}_i^{(m)} - \dot{u}_i\right)^2 + \frac{\phi_1\phi_2^2}{6}\dot{\zeta}^2\beta_1, \quad (3)$$

where $b_m = v_m\phi_{m0}^2\eta_m/\kappa_m$ ($m = 1, 2$) is the viscous resistance coefficients and $\kappa_m$ is the permeability. $\eta_m$ is the fluid viscosity and $v_m$ is volume fractions of solid inclusion and background material in the REV. Here, $\kappa_1 = \langle\kappa\rangle_{in}$ and $\kappa_2 = \langle\kappa\rangle_{host}$ are used. The term related to LFIF is $\beta_1 = v_1^{2/3}\phi_{10}a_{20}^2[b_1 + b_2(1 - a_{10}/a_{20})]$.

Regarding the Euler–Lagrange equations with dissipation function, the dynamic equations are

$$\rho_{00}\ddot{\mathbf{u}} + \sum_{m=1}^{2}[\rho_{0m}\ddot{\mathbf{U}}^{(m)} + b_m(\dot{\mathbf{u}} - \dot{\mathbf{U}}^{(m)})] = \nabla\cdot\boldsymbol{\sigma}, \quad (4)$$

$$\rho_{01}\ddot{\mathbf{u}} + \rho_{11}\ddot{\mathbf{U}}^{(1)} + b_1(\dot{\mathbf{U}}^{(1)} - \dot{\mathbf{u}}) = \nabla s_1, \quad (5)$$

$$\rho_{02}\ddot{\mathbf{u}} + \rho_{22}\ddot{\mathbf{U}}^{(2)} + b_2(\dot{\mathbf{U}}^{(2)} - \dot{\mathbf{u}}) = \nabla s_2, \quad (6)$$

$$-(\chi_1\ddot{\zeta}_1 + \beta_1\dot{\zeta}_1)\frac{\varphi_1\varphi_2^2}{3} = \varphi_2 s_1 - \varphi_1 s_2, \quad (7)$$

where $\boldsymbol{\sigma} = \frac{\partial W}{\partial \mathbf{e}}$, $s_1 = \frac{\partial W}{\partial \xi_1}$, $s_2 = \frac{\partial W}{\partial \xi_2}$. Here, $\mathbf{e}$ is the strain of solid frame. The scaler $s_1, s_2$ are related to fluid pressure by $s_1 = -\phi p_1$ and $s_2 = -\phi p_2$. $p_1$ and $p_2$ are the pressure in inner fluid pocket and outside fluid patch. The viscous resistance coefficients can also be written as $b_m = \rho_{f_m}\phi_m\omega_c^{(m)}$ ($m=1, 2$). Here, $\omega_c^{(m)} = \frac{\mu_m\phi_{m0}}{\rho_{f_m}\kappa_m}$ is the characteristic frequency for immiscible fluid patches.

### B. Wave equations of Biot-patchy-squirt (BIPS) model

Consider a cubic REV in the partially saturated double-porosity medium, which has a size larger enough than pores and patches but smaller enough than the macroscopic scale (such as the wavelength). BISQ model provides a scenario of transverse fluid flow in a representative cylinder with its axis parallel to the direction of wave propagation. For single-porosity case, 2D mass conservation equations in $x$ and $r$ directions have been given by Dvorkin and Nur.[61] Here, squirt flow is extended to patchy-saturation in double-porosity rocks.

For the case of uniaxial deformation, the axial dynamic equations are

$$\rho_{00}\ddot{u}_x + \rho_{01}\ddot{U}_x^{(1)} + \rho_{02}\ddot{U}_x^{(2)} + b_1\left(\dot{u}_x - \dot{U}_x^{(1)}\right) + b_2\left(\dot{u}_x - \dot{U}_x^{(2)}\right)$$
$$= \bar{M}\frac{d^2 u_x}{dx^2} - \gamma_1\frac{dp_1}{dx} - \gamma_2\frac{dp_2}{dx},$$

$$\rho_{01}\ddot{u}_x + \rho_{11}\ddot{U}_x^{(1)} + b_1\left(\dot{U}_x^{(1)} - \dot{u}_x\right) = -\phi_1\frac{\partial p_1}{\partial x}, \quad (8)$$

$$\rho_{02}\ddot{u}_x + \rho_{22}\ddot{U}_x^{(2)} + b_2\left(\dot{U}_x^{(2)} - \dot{u}_x\right) = -\phi_2\frac{\partial p_2}{\partial x},$$

$$-(\chi_1\ddot{\zeta}_x + \beta_1\dot{\zeta}_x)\frac{\phi_1\phi_2^2}{3} = \phi_1\phi_2(p_2 - p_1),$$

where $\zeta_x = \frac{dZ_x}{dx}$ and $Z_x$ is the x-direction component of the local fluid flow between different fluid patches. $\bar{M} = \frac{M_1 M_2}{\nu_1 M_2 + \nu_2 M_1}$, $M_m = 2G_m \frac{1-\nu_m}{1-2\nu_m}$, $m = 1, 2$. Here, $G_m$ and $\nu_m$ are shear modulus and Poisson's ratio of the solid inclusion ($m = 1$) and surrounding background material ($m = 2$).

The radial dynamic equations are

$$\frac{\rho_{f_1}}{2}\phi_1\left(1+\frac{1}{\phi_{10}}\right)\ddot{U}_r^{(1)} + b_1 \dot{U}_r^{(1)} = -\phi_1 \frac{\partial p_1}{\partial r},$$
$$\frac{\rho_{f_2}}{2}\phi_2\left(1+\frac{1}{\phi_{20}}\right)\ddot{U}_r^{(2)} + b_2 \dot{U}_r^{(2)} = -\phi_2 \frac{\partial p_2}{\partial r}. \quad (9)$$

The conservation equations in a partially saturated double-porosity medium are given as (Appendix A)

$$\frac{\phi_1}{\rho_{f_1}}\frac{\partial \rho_{f_1}}{\partial t} + \frac{\partial \phi_1}{\partial t} + \phi_1 \frac{\partial^2 \left(U_x^{(1)} - \phi_2 Z_x - u_x\right)}{\partial t \partial x}$$
$$+ \phi_1 \left(\frac{\partial^2 \left(U_r^{(1)}\right)}{\partial t \partial r} + \frac{1}{r}\frac{\partial \left(U_r^{(1)}\right)}{\partial t}\right) = 0,$$
$$\frac{\phi_2}{\rho_{f_2}}\frac{\partial \rho_{f_2}}{\partial t} + \frac{\partial \phi_2}{\partial t} + \phi_2 \frac{\partial^2 \left(U_x^{(2)} + \phi_1 Z_x - u_x\right)}{\partial t \partial x}$$
$$+ \phi_2 \left(\frac{\partial^2 \left(U_r^{(2)}\right)}{\partial t \partial r} + \frac{1}{r}\frac{\partial \left(U_r^{(2)}\right)}{\partial t}\right) = 0. \quad (10)$$

In the case of plane wave propagation, the displacements and pressure are expressed as $[u_x, U_x^{(1)}, U_x^{(2)}, U_r^{(1)}, U_r^{(2)}, Z_x, p_1, p_2] = [C_1, C_2^{(1)}, C_2^{(2)}, C_{r0}^{(1)}, C_{r0}^{(2)}, C_3, p_{01}, p_{02}] \cdot e^{i(\omega t - \mathbf{k} \cdot \mathbf{x})}$.

Combing (8), (9), and (10), the ordinary differential equations of $p_{01}$ and $p_{02}$ are

$$\frac{\partial^2 p_{01}}{\partial r^2} + \frac{1}{r}\frac{\partial p_{01}}{\partial r} + \frac{\rho_{f_1}\omega^2}{F_1}\left(\frac{1+1/\phi_{10}}{2} - i\frac{\omega_c^{(1)}}{\omega}\right)p_{01}$$
$$- ik\rho_{f_1}\omega^2 \left(C_2^{(1)} - \phi_2 C_3 + \frac{\gamma_1}{\phi_1}C_1\right)\left(\frac{1+1/\phi_{10}}{2} - i\frac{\omega_c^{(1)}}{\omega}\right) = 0, \quad (11)$$

$$\frac{\partial^2 p_{02}}{\partial r^2} + \frac{1}{r}\frac{\partial p_{02}}{\partial r} + \frac{\rho_{f_2}\omega^2}{F_2}\left(\frac{1+1/\phi_{20}}{2} - i\frac{\omega_c^{(2)}}{\omega}\right)p_{02}$$
$$- ik\rho_{f_2}\omega^2 \left(C_2^{(2)} + \phi_1 C_3 + \frac{\gamma_2}{\phi_2}C_1\right)\left(\frac{1+1/\phi_{20}}{2} - i\frac{\omega_c^{(2)}}{\omega}\right) = 0.$$

This is the Bessel differential equation of pressure $p_{0m}$ ($m = 1, 2$). Adopting pressure boundary condition $p_{01} = p_{01}^*$, $p_{02} = p_{02}^*$ at pore wall $r = R$, pressures are obtained as

$$p_{01} = ikF_1\left(C_2^{(1)} - \phi_2 C_3 + \frac{\gamma_1}{\phi_1}C_1\right)\left(1 - \frac{J_0(\lambda_1 r)}{J_0(\lambda_1 R)}\right) + p_{01}^* \frac{J_0(\lambda_1 r)}{J_0(\lambda_1 R)},$$
$$p_{02} = ikF_2\left(C_2^{(2)} + \phi_1 C_3 + \frac{\gamma_2}{\phi_2}C_1\right)\left(1 - \frac{J_0(\lambda_2 r)}{J_0(\lambda_2 R)}\right) + p_{02}^* \frac{J_0(\lambda_2 r)}{J_0(\lambda_2 R)}, \quad (12)$$

where $F_1 = \frac{K_{f_1}(\phi_1 + \phi_2 \beta)H_1}{K_{f_1}+(\phi_1+\phi_2\beta)H_1}$, $F_2 = \frac{K_{f_2}(\phi_2+\phi_1/\beta)H_2}{K_{f_2}+(\phi_2+\phi_1/\beta)H_2}$. $J_0(\lambda r)$ is the zero order Bessel function. $p_{01}^*$ and $p_{02}^*$ are pressure values at the boundary. The coefficient $\lambda_1, \lambda_2$ are defined as $\lambda_1^2 = \frac{\rho_{f_1}\omega^2}{F_1}\left(\frac{1+1/\varphi_{10}}{2} - i\frac{\omega_c^{(1)}}{\omega}\right)$ and $\lambda_2^2 = \frac{\rho_{f_2}\omega^2}{F_2}\left(\frac{1+1/\varphi_{20}}{2} - i\frac{\omega_c^{(2)}}{\omega}\right)$.

The squirt flow model has been developed for partially saturated rocks where cross-flow exists between fluid pore and gas pockets,[61] where the pressure boundary condition at pore wall $r = R$ is set to zero.

### C. P-wave dispersion and attenuation

The dynamic equations of Biot-patchy-squirt model relate displacements $U_x, u_x, Z_x$ to fluid pressure $p$. The pressure boundary condition plays an essential role in determining the bulk modulus and velocity of porous media. Zero-pressure boundary condition on the surface of the representative cylinder has been applied for squirt flow.[61] Such boundary condition leads to a lower P-wave velocity than Biot's theory and Gassmann equation at low frequency limit. Non-zero pressure condition has been employed by using pore pressure increment at the surface of unrelaxed fraction of pore space to model squirt flow in fully saturated rocks.[3] The pressure increment is related to confining pressure. In this way, the undrained pore rocks have relatively high frame moduli. And the predicted P-wave velocity at low frequency limit is identical to Gassmann's equation.

In the case of zero pressure boundary condition ($p_m^* = 0$, m = 1, 2), the average flow pressures $\bar{p}_1$ and $\bar{p}_2$ in the cylinder channel are determined,

$$\bar{p}_1 = \frac{1}{\pi R^2}\int_0^R 2\pi r p_1(x,r,t)dr = -\tilde{F}_1\left(\frac{\partial U_x^{(1)}}{\partial x} - \phi_2 \frac{\partial Z_x}{\partial x} + \frac{\gamma_1}{\phi_1}\frac{\partial u_x}{\partial x}\right),$$
$$\bar{p}_2 = \frac{1}{\pi R^2}\int_0^R 2\pi r p_2(x,r,t)dr = -\tilde{F}_2\left(\frac{\partial U_x^{(2)}}{\partial x} + \phi_1 \frac{\partial Z_x}{\partial x} + \frac{\gamma_2}{\phi_2}\frac{\partial u_x}{\partial x}\right), \quad (13)$$

where $\tilde{F}_m = F_m\left(1 - \frac{2J_1(\lambda_m R)}{\lambda_m R J_0(\lambda_m R)}\right)$. In the case that the pressure $p$ in dynamic equations (8) is approximated by its average value $\bar{p}$, the wave equations in $\omega$ and $k$ space are

$$\begin{bmatrix} a_{11}Y + b_{11} & a_{12}Y + b_{12} & a_{13}Y + b_{13} \\ a_{21}Y + b_{21} & a_{22}Y + b_{22} & a_{23}Y + b_{23} \\ a_{31}Y + b_{31} & a_{32}Y + b_{32} & a_{33}Y + b_{33} \end{bmatrix} \begin{Bmatrix} C_1 \\ C_2^{(1)} \\ C_2^{(2)} \end{Bmatrix} = 0, \quad (14)$$

where $Y = \left(\frac{k}{\omega}\right)^2$, $m = 1, 2$. The wave equations have nonzero solutions for constant $C_1, C_2^{(1)}, C_2^{(2)}$ only if the determinant is zero, which provides an equation

$$AY^3 + BY^2 + CY + D = 0. \quad (15)$$

The coefficients $a_{ij}, b_{ij}$ and $A, B, C, D$ are given in Appendix B. The complex and phase velocities of P-wave are defined as $v = \omega/k$, $v_P = \mathbf{Re}(v)$. The P-wave attenuation is obtained as $Q^{-1} = \mathbf{Im}(v^2)/\mathbf{Re}(v^2)$.

## III. DISCUSSIONS

### A. BIPS reduces to BISQ in the condition of single-porosity and full-saturation

When the porous medium is composed of single porosity and fully saturated by one fluid, without loss of generality, one can have $v_1 \to 1$ and $v_2 = 0$, $b_2 = 0$, $\rho_{02} = \rho_{22} = 0$. In such case, the coefficients $a_{ij}$, $b_{ij}$ are

$$a_{11} = M + \tilde{F}_1 \frac{(\alpha - \phi_{10})^2}{\phi_{10}}, \quad (16)$$

$$a_{12} = a_{21} = \tilde{F}_1(\alpha - \phi_{10}), \quad (17)$$

$$a_{13} = a_{31} = 0, \quad (18)$$

$$a_{22} = \tilde{F}_1 \phi_{10}, \quad (19)$$

$$a_{23} = a_{32} = 0, \quad (20)$$

$$a_{33} = 0, \quad (21)$$

$$b_{11} = -\rho_{00} + \frac{ib_1}{\omega}, \quad (22)$$

$$b_{12} = b_{21} = -\rho_{01} - \frac{ib_1}{\omega}, \quad (23)$$

$$b_{13} = b_{31} = 0, \quad (24)$$

$$b_{22} = -\rho_{11} + \frac{ib_1}{\omega}, \quad (25)$$

$$b_{23} = b_{32} = 0, \quad (26)$$

$$b_{33} = 0. \quad (27)$$

Applying a linear substitution to the reduced plane wave equations, wave equations in $\omega$ and $k$ space are

$$\begin{bmatrix} \frac{1}{\rho_1}\left(M + \frac{\alpha\gamma\tilde{F}}{\phi}\right)Y - \frac{\rho_0}{\rho_1} & \frac{\alpha\tilde{F}}{\rho_1}Y - 1 \\ \frac{\gamma\tilde{F}}{\rho_1}Y - \frac{\rho_{01}}{\rho_1} - \frac{i\omega_c}{\omega} & \frac{\phi\tilde{F}}{\rho_1}Y - 1 + \frac{\rho_{01}}{\rho_2} + i\frac{\omega_c}{\omega} \end{bmatrix} \begin{Bmatrix} C_1 \\ C_2^{(1)} \end{Bmatrix} = 0. \quad (28)$$

Here, parameter substitutions are used as $\tilde{F}_1 = \tilde{F}$, $\phi_{10} = \phi$, $b_1 = \phi_f \phi \frac{i\omega_c}{\omega}$.

Equation (28) is exactly the same as in BISQ model given by Dvorkin and Nur,[61] except for some trivial differences. In BISQ model, $\rho_1$ and $\rho_2$ are densities related to solid and fluid, respectively, while here the densities are $\rho_0$ and $\rho_1$. The couple density coefficient is $-\rho_a$ in BISQ model, but it is $\rho_{01}$ here. The characteristic frequency term is $-i\frac{\omega_c}{\omega}$ in BISQ model, while here it is $i\frac{\omega_c}{\omega}$ in BIPS model. The opposite sign comes from that fact that $e^{i(\omega t - \mathbf{k} \cdot \mathbf{x})}$ is used in BIPS model in the displacements and pressures, while $e^{i(lx - \omega t)}$ is used in BISQ model.

### B. BIPS reduces to Biot's theory at infinity frequency

As shown in Sec. III A, BIPS is reduced to BISQ when the porous medium is single-porosity and fully saturated by one fluid. In such case, it is easy to show that Biot's theory is obtained from BIPS at infinity frequency. For uniaxial deformation in the $x$-direction, the Biot's equations can be rewritten as

$$(A + 2N)\nabla^2 u_x + Q\nabla^2 U_x = \frac{\partial^2}{\partial t^2}(\rho_{11} u_x + \rho_{12} U_x) + b\frac{\partial}{\partial t}(u_x - U_x),$$

$$Q\nabla^2 u_x + R\nabla^2 U_x = \frac{\partial^2}{\partial t^2}(\rho_{12} u_x + \rho_{22} U_x) - b\frac{\partial}{\partial t}(u_x - U_x). \quad (29)$$

In the case of $u_x = u_{x0} e^{i(\omega t - \mathbf{k} \cdot \mathbf{x})}$, $U_x = U_{x0} e^{i(\omega t - \mathbf{k} \cdot \mathbf{x})}$, a linear substitution of (29) gives

$$[(A + 2N + Q)Y - \rho_1]u_x + [(Q + R)Y - \rho_2]U_x = 0,$$

$$\left(QY + \rho_a + \frac{ib}{\omega}\right)u_x + \left(RY - \rho_2 - \rho_a - \frac{ib}{\omega}\right)U_x = 0, \quad (30)$$

where $b = \mu \phi^2 / \kappa$ and $Y = (k/\omega)^2$. Here, densities satisfy $\rho_1 = \rho_{11} + \rho_{12}$, $\rho_2 = \rho_{22} + \rho_{12}$ and $\rho_a = -\rho_{12}$ in the derivations. Compared with the Biot's equations given in Appendixes A and B of the work by Dvorkin and Nur,[61]

$$C_1\left[Y\frac{1}{\rho_2}\left(M + \frac{\alpha\gamma F}{\phi}\right) - \frac{\rho_1}{\rho_2}\right] + C_2\left(Y\frac{\alpha F}{\rho_2} - 1\right) = 0,$$

$$C_1\left(Y\frac{\gamma F}{\rho_2} + \frac{\rho_a}{\rho_2} + i\frac{\omega_c}{\omega}\right) + C_2\left(Y\frac{\phi F}{\rho_2} - 1 - \frac{\rho_a}{\rho_2} - i\frac{\omega_c}{\omega}\right) = 0, \quad (31)$$

where $\omega_c = \frac{\mu\phi}{\kappa\rho_f}$, one can easily recognize that Eqs. (30) and (31) are exactly the same. The term in BISQ and Biot's equations is equivalent as

$$A + 2N + Q = M + \frac{\alpha\gamma F}{\phi}. \quad (32)$$

Dvorkin and Nur[61] have shown that as $\omega \to \infty$, $\tilde{F} \to F$. In such case, the P-wave velocity of BISQ model is identical to Biot's theory. As a consequence, the BIPS model in single-porosity fully saturated media is degraded to BISQ model at high frequency limit, and hence is reduced to Biot's theory.

### C. BIPS reduces to patchy-saturation model at high frequency limit

For double-porosity and partially saturated medium, the BIPS model incorporates both local fluid flow and squirt flow mechanisms. As the frequency approaches infinity, the coefficients $\tilde{F}_m$ satisfy

$$\tilde{F}_m = F_m\left(1 - \frac{2J_1(\lambda_m R)}{\lambda_m R J_0(\lambda_m R)}\right) \to F_m. \quad (33)$$

The physical meaning, as suggested by Dvorkin and Nur,[61] is clear that the fluid at high frequency limit is in unrelaxed mode and cannot be squeezed out of the pore space.

Here, it will be shown that BIPS model degenerates to double-porosity dual-fluid patch model[14] as the squirt flow is canceled out at high frequency. For uniaxial deformation in the x-direction, wave equations of patchy-saturation model can be written as[14]

$$[(A + 2N + Q_1 + Q_2)Y - \rho_{00} - \rho_{01} - \rho_{02}]u_x$$
$$+ \sum_{m=1}^{2}[(Q_m + R_m)Y - \rho_{0m} - \rho_{mm}]U_x^{(m)}$$
$$+ [Q_2\phi_1 - Q_1\phi_2 + R_2\phi_1 - R_1\phi_2]YZ_x = 0, \quad (34)$$

$$\left(Q_1 Y - \rho_{01} - \frac{ib}{\omega}\right)u_x + \left(R_1 Y - \rho_{11} + \frac{ib_1}{\omega}\right)U_x^{(1)} - R_1 Y \phi_1 Z_x = 0, \tag{35}$$

$$\left(Q_2 Y - \rho_{02} - \frac{ib}{\omega}\right)u_x + \left(R_2 Y - \rho_{22} + \frac{ib_2}{\omega}\right)U_x^{(1)} - R_2 Y \phi_2 Z_x = 0, \tag{36}$$

$$(\phi_2 Q_1 - \phi_1 Q_2)u_x + \phi_2 R_1 U_x^{(1)} - \phi_1 R_2 U_x^{(2)}$$
$$+ \left[\phi_2^2 R_1 - \phi_1^2 R_2 - (\chi_1^* \omega^2 - i\omega \beta_1^*)\frac{\phi_{10}\phi_2^2}{3}r_{10}^2\right]Z_x = 0, \tag{37}$$

where $Y = (k/\omega)^2$, $\chi_1^* = \left(\rho_{f_1} + \rho_{f_2}\frac{\phi_{10}}{\phi_{20}}[1 - \nu_1^{1/3}]\right)\nu_1$, and $\beta_1^* = \left(\frac{\eta_1}{\kappa_{10}} + \frac{\eta_2}{\kappa_{20}}(1 - \nu_1^{1/3})\right)\phi_{10}\nu_1$.

By substituting (13) into (8), we have the wave equations of BIPS model. After a linear substitution, the equations are written as

$$\left[\left(\bar{M} + \frac{\tilde{F}_1 \gamma_1^2}{\phi_1} + \frac{\tilde{F}_2 \gamma_2^2}{\phi_2} + \tilde{F}_1 \gamma_1 + \tilde{F}_2 \gamma_2\right)Y - \rho_{00} - \rho_{01} - \rho_{02}\right]u_x$$
$$+ \sum_{m=1}^{2}(\tilde{F}_m \gamma_m Y - \rho_{0m} - \rho_{mm})U_x^{(m)}$$
$$+ (\tilde{F}_2 \gamma_2 \phi_1 - \tilde{F}_1 \gamma_1 \phi_2 + \tilde{F}_2 \phi_2 \phi_1 - \tilde{F}_1 \phi_1 \phi_2)Y Z_x = 0, \tag{38}$$

$$\left[\gamma_1 \tilde{F}_1 Y - \rho_{01} - \frac{ib_1}{\omega}\right]u_x + \left[\phi_1 \tilde{F}_1 Y - \rho_{11} + \frac{ib_1}{\omega}\right]U_x^{(1)}$$
$$- \phi_1 \phi_2 \tilde{F}_1 Y Z_x = 0, \tag{39}$$

$$\left[\gamma_2 \tilde{F}_2 Y - \rho_{02} - \frac{ib_2}{\omega}\right]u_x + \left[\phi_2 \tilde{F}_2 Y - \rho_{22} + \frac{ib_2}{\omega}\right]U_x^{(2)}$$
$$+ \phi_1 \phi_2 \tilde{F}_2 Y Z_x = 0, \tag{40}$$

$$(\phi_1 \gamma_2 \tilde{F}_2 - \phi_2 \gamma_1 \tilde{F}_1)u_x - \phi_1 \phi_2 \tilde{F}_1 U_x^{(1)} + \phi_1 \phi_2 \tilde{F}_2 U_x^{(2)}$$
$$+ \left[(\chi_1 \omega^2 - i\beta_1 \omega)\frac{\phi_1 \phi_2^2}{3} + \phi_1 \phi_2^2 \tilde{F}_1 + \phi_1^2 \phi_2 \tilde{F}_2\right]Z_x = 0. \tag{41}$$

It is clear that when $\omega \to \infty$, there exists $\tilde{F}_m \to F_m$. The details of the mathematical derivation will not be given here, but it is not difficult to derive following relations for BIPS and patchy models:

$$A + 2N = \bar{M} + \frac{F_1 \gamma_1^2}{\phi_1} + \frac{F_2 \gamma_2^2}{\phi_2}, \tag{42}$$

$$Q_1 = F_1 \gamma_1, \quad Q_2 = F_2 \gamma_2, \tag{43}$$

$$R_1 = F_1 \phi_1, \quad R_2 = F_2 \phi_2. \tag{44}$$

The terms before $u_x$, $U_x^{(1)}$, $U_x^{(2)}$, and $Z_x$ in (37) and (41) are equivalent to each other. As a consequence, at the high frequency limit, the BIPS model is reduced to patchy-saturation model. Please note that the patch size $r_{10}$ appear in TLP model.[14] However, in BIPS model, the patch size obeys a random distribution, rather than a single predefined value. In such a way, BIPS model will not explicitly depend on patch size $r_{10}$.

## IV. NUMERICAL EXAMPLES

The dispersion and attenuation of P-wave propagating in porous media depend on fluid density, modulus, viscosity, and skeleton properties (such as bulk modulus, density and permeability). In BISQ model, the characteristic squirt length is another important parameter. The effects of permeability and characteristic squirt length on P-wave velocity have been studied by Dvorkin and Nur.[61] In BIPS model, P-wave velocity not only depends on the rock physics parameters of the double-porosity matrix, but also the parameters of fluid patch. In addition, the fluid patch volume fraction in BIPS model is an extra parameter as important as characteristic squirt length in BISQ model. In this section, the effects of permeability, viscosity, patch volume fractions, and characteristic squirt length on P-wave dispersion and attenuation are investigated.

### A. The effects of frequency and fluid type

First, the effects of frequency and fluid type are to be investigated. The rock physics parameters are from the numerical examples in the work of Dvorkin and Nur.[61] The porosity is 0.15. The bulk modulus and Poisson's ratio of matrix are 16 GPa and 0.15. The modulus and density of solid phase are $K_s = 38$ GPa and 2650 kg/m$^3$. The velocity vs frequency curves are calculated by BIPS model and compared with Biot's theory, Gassmann's equation, BISQ, and TLP model (Fig. 2).

In the first case, the fluid in the inner pocket is gas (modulus $K_{gas} = 0.142$ MPa, density $\rho_{gas} = 1.2$ kg/m$^3$, viscosity $\eta_{gas} = 18.6$ $\mu$Pa s). The surrounding fluid is water (modulus $K_{water} = 2.2$ GPa, density $\rho_{water} = 1000$ kg/m$^3$, viscosity $\eta_{water} = 1.0$ mPa s). The volume fraction of inner gas pocket is set to 1, i.e., the matrix is fully saturated by gas.

The results in Fig. 2 give following features. (1) At high frequency limit, the squirt flow effect is switched off, i.e., $\tilde{F} \to F$, the velocit dispersion [Fig. 2(a)] and attenuation [Fig. 2(b)] of BIPS model match very well with that of Biot's theory. (2) At low frequency, the velocities of BIPS and BISQ models are a little bit lower than Biot's theory [Fig. 2(a)], but the difference is extremely tiny in full gas saturation case (around 3968.57–3968.59 m/s). (3) The predicted dispersion and attenuation by TLP model overlap with Biot's theory very well. The reason is that TLP model incorporates Biot global flow mechanism and local fluid flow at fluid interfaces simultaneously. In the fully gas saturation case, there is only one fluid and the LFIF effect at patch interface disappears. Thus TLP model degenerates to Biot's theory. (4) It is clear that BIPS and BISQ models have two attenuation peaks [Fig. 2(b)]. The peak at high frequency (between 1 and 100 MHz) is related to Biot mechanism, which overlaps perfectly with the curve of Biot's theory. The peak at lower frequency (between 100 Hz and 10 kHz) is caused by the squirt effect, which has been taken into consideration in BIPS and BISQ models.

In the second case, the inner pocket is filled with water and the surrounding fluid is gas. The volume fraction of water pocket is set to 1, i.e., the matrix is fully saturated by water. The velocity dispersion and attenuation are shown in Fig. 3. The water-saturated case shows consistent behaviors as observed in the gas-saturated case. In addition, it is worth noting that squirt flow effect is apparently enhanced in water-saturation case. (1) The velocities of BIPS and BISQ at low frequency are substantially lower than that of Biot's theory. The velocity difference has been attributed to the constant pressure boundary condition.[61] (2) The two attenuation peaks of BIPS and BISQ models, i.e., the squirt peak at low frequency and Biot peak at high frequency, move toward each other and combine into a single larger peak. The Biot peaks of TLP model and Biot's theory are still there, but shift toward lower frequency.

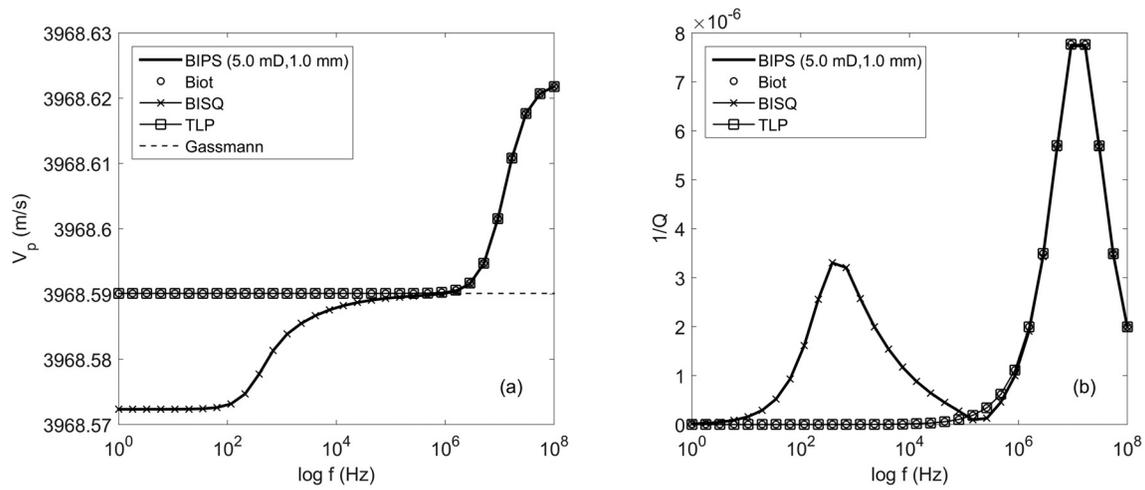

**FIG. 2.** Prediction of P-wave velocity dispersion (a) and attenuation (b) by BIPS, BISQ, TLP, Biot, and Gassmann methods. The volume fraction of inner fluid pocket (gas) is 1. The overall matrix has uniform porosity.

It should be noted that the movement of attenuation peak along frequency axis depends not only on fluid mobility (viscosity and permeability), but also on fluid density and bulk modulus. It has been observed that as the viscosity of pore fluid increases, or matrix permeability decreases, the attenuation peak of squirt mechanism shifts toward lower frequency.[71] The TLP peaks move in the same direction as squirt model. On the other hand, Biot peak shifts in the opposite direction. Here, we can find that as gas-saturation transits to water-saturation, the fluid viscosity increases from $\eta_{gas} = 18.6$ $\mu$Pa s to $\eta_{water} = 1.0$ mPa s. The attenuation peak is supposed to shift toward low frequency since the fluid mobility decreases as viscosity increases. Surprisingly, the squirt peak and TLP peak shift to a higher frequency region. The reason for this discrepancy comes from the fact that the fluid modulus increases from $K_{gas} = 0.142$ MPa to $K_{water} = 2.2$ GPa and density increases from $\rho_{gas} = 1.2$ kg/m$^3$ to $\rho_{water} = 1000$ kg/m$^3$.

In summary, in the single-fluid saturation case, the patch effect disappears and BIPS model is the same as BISQ model. BIPS and BISQ models contain both squirt and Biot attenuation peaks. When the fluid changes from gas to water, the squirt effect is substantially enhanced. In addition, attenuation peaks caused by squirt and Biot mechanisms move toward each other and combine into a large peak when the matrix is fully saturated by water. When fluid mobility decreases, the attenuation peak of Biot's theory shifts toward higher frequency and that of squirt flow model shifts toward lower frequency, which explains the fusion of two separate peaks.

### B. The effects of patch saturation and permeability

In this section, the effects of patch saturation and permeability on P-wave velocity dispersion and attenuation are to be investigated.

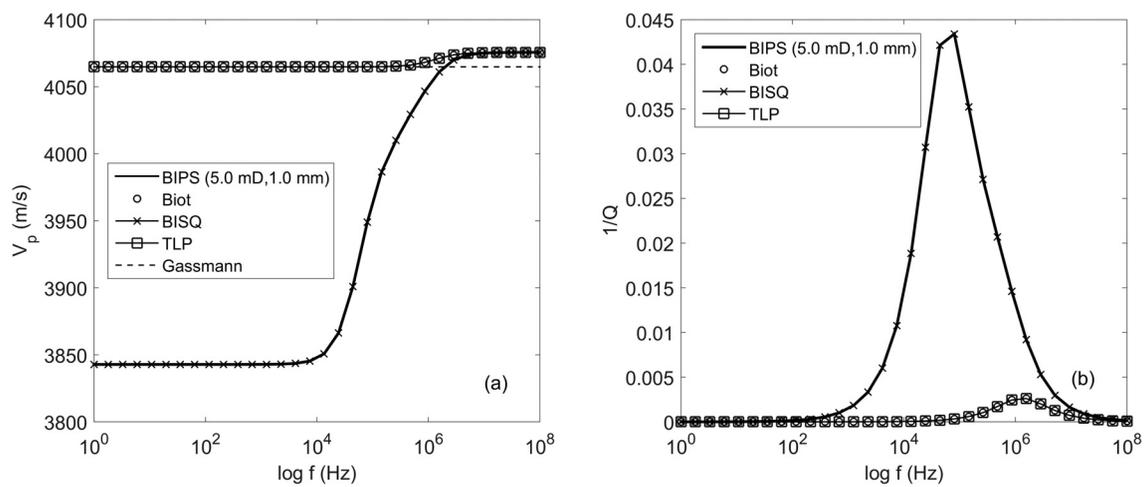

**FIG. 3.** Prediction of P-wave velocity dispersion (a) and attenuation (b) by BIPS, BISQ, TLP, Biot, and Gassmann methods. The volume fraction of inner fluid pocket (water) is 1. The overall matrix has uniform porosity.

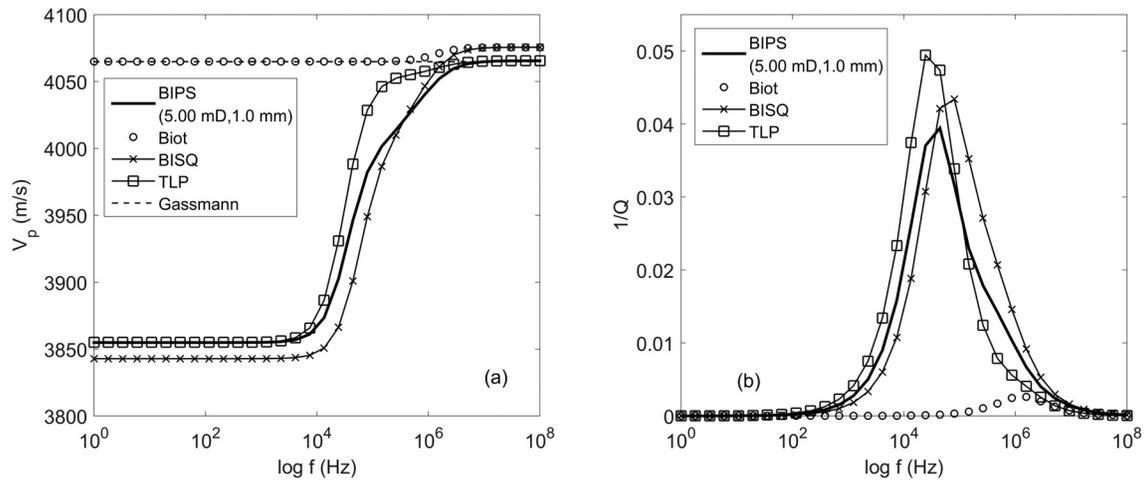

**FIG. 4.** Prediction of P-wave velocity dispersion (a) and attenuation (b) by BIPS, BISQ, TLP, Biot and Gassmann methods. The volume fraction of inner fluid pocket (water) is 0.9. The permeability is 5 mD. The squirt characteristic length is 1 mm.

The inner fluid pockets are filled with water in BIPS and TLP models. The pocket volume fraction is set to 0.9 (Fig. 4) and 0.3 (Fig. 5), respectively, to check the influence of patchy-saturation. In predicting velocity by Biot's theory and BISQ model, the matrix is saturated by water. The rock physics parameters are the same as in Sec. IV A. BISQ characteristic radius is set to 1 mm here.

To find out the underlying reasons causing the velocity difference among BIPS, BISQ, and TLP models, the P-wave dispersion/attenuation under different mechanisms have been calculated (Figs. 4 and 5). The results show the following apparent features. (1) In partial saturation case, the TLP velocity is much lower than Biot and Gassmann velocity at low frequency, rather than overlap with Biot's theory as in full-saturation case. The reason for the velocity decrease is that an interface emerges between the inner water pocket and surround gas patch. The patch effect in TLP model arises from the LFIF at the interface and causes a decrease in velocity remarkably. (2) As one can see in Figs. 4(b) and 5(b), BIPS, BISQ, TLP, and Biot attenuation peaks have difference heights and locations. As volume fraction of inner water pocket decrease from 0.9 to 0.3 [Fig. 5(b)], the BIPS and TLP peaks are dwarfed, while BISQ and Biot peaks do not change. The physical meaning of this sharp change in peak height is that as the pocket fluid decreases, the kinetic energy dissipation caused by patchy mechanism in BIPS and TLP models is weakened. (3) It is clear that the BIPS peak has a bump near the location of Biot peak [Figs. 4(b) and 5(b)], which indicates that BIPS model includes the effect of Biot mechanism. (4) An interesting observation is that the BIPS peak is lower than the TLP peak [Figs. 4(b) and 5(b)]. In the current partial-saturation case, BIPS model includes Biot, patchy, and squirt mechanism simultaneously. A reasonable explanation is that patchy-saturation effect can be weakened by squirt flow effect. The LFIF dissipation at the patch interface is to be partially canceled by the existence of squirt flow, which has been suggested by Johnson.[6]

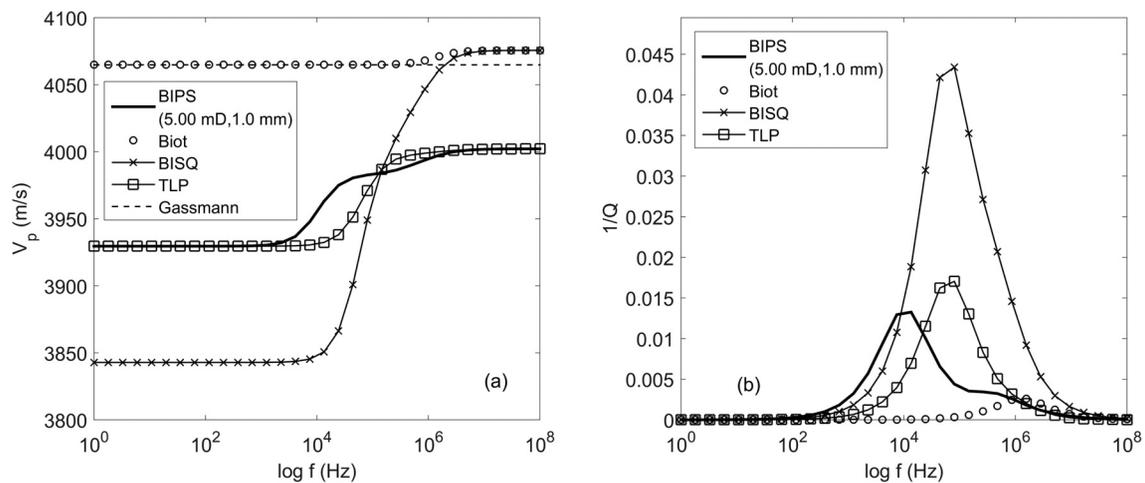

**FIG. 5.** Prediction of P-wave velocity dispersion (a) and attenuation (b) by BIPS, BISQ, TLP, Biot, and Gassmann methods. The volume fraction of inner fluid pocket (water) is 0.3. The permeability is 5 mD. The squirt characteristic length is 1 mm.

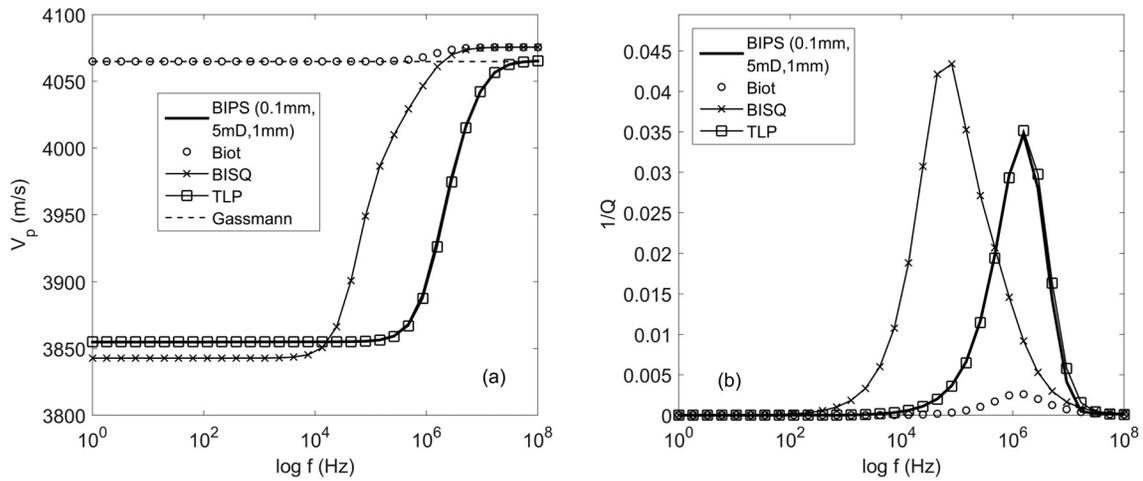

**FIG. 6.** Prediction of P-wave velocity dispersion (a) and attenuation (b) by BIPS, BISQ, TLP, Biot, and Gassmann methods. The volume fraction of inner fluid pocket (water) is 0.9. The permeability is 5 mD. The squirt characteristic length is 1 mm. The whole patch size is 0.1 mm.

As shown above, LFIF and squirt flow are identified as separate mechanisms by comprehensive analysis of frequency, fluid type, and fluid pocket size. Moreover, the essential difference and coupling effects between patchy and squirt effects are in the form of attenuation peak variation as the whole concentric patch size changes. To facilitate the understanding of the relation among BIPS, BISQ, and TLP model, the whole patch (including the inner pocket and surrounding fluid patch) size is set to 0.1 and 5 mm, respectively. The results are shown in Figs. 6 and 7.

As shown in Fig. 6(b), the attenuation peaks of BIPS and TLP, which are located at high frequency region (around 1 MHz), are at the right hand side of BISQ peak. Moreover, BIPS and TLP peaks overlap with each other quite well, suggesting that BIPS is reduced to TLP at high frequency limit, which has been theoretically analyzed in Discussions section.

As the whole patch size increases to 5 mm, the TLP peak has a remarkable shift to lower frequency (around 1 kHz) [Fig. 7(b)]. Correspondingly, the P-wave velocity increases lot for the same frequency [Fig. 7(a)]. The physical explanation which underlines this velocity enhancement is that larger patch size, which indicates larger fluid heterogeneity, requires much longer time for the fluid to relax. The ratio between mesoscopic fluid patch size and diffusion length is of great importance for patchy-saturation mechanism. It describes the velocity dispersion caused by LFIF between inner fluid pocket and surrounding fluid patch. Since the fluid does not have enough time to relax in the larger patches, the porous rock is actually in an unrelaxed state.

Surprisingly, there is no BIPS peak at low frequency. As demonstrated above, BIPS is a general model that incorporates the three important mechanisms: Biot mechanism, patchy-saturation mechanism, and squirt flow mechanism. The patchy-saturation, which has also been incorporated in TLP model, causes the attenuation peak shift toward low frequency. However, the effect of patchy-saturation, i.e., a shift of attenuation peak to low frequency, does not appear in BIPS model. More importantly, the BIPS peak overlaps the BISQ peak pretty well, except for a reduction in peak height. A reasonable interpretation is that squirt flow mechanism suppresses the patchy saturation effects to a great extent and trim the patchy-saturation peak completely, which has been demonstrated in Figs. 4(b) and 5(b) and also in other works.[6,72]

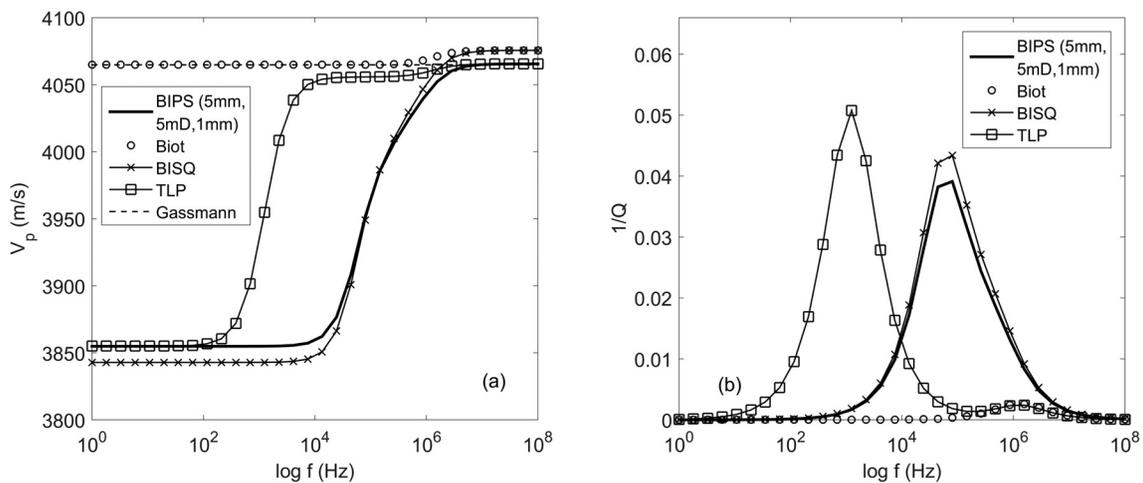

**FIG. 7.** Prediction of P-wave velocity dispersion (a) and attenuation (b) by BIPS, BISQ, TLP, Biot, and Gassmann methods. The volume fraction of inner fluid pocket (water) is 0.9. The permeability is 5 mD. The squirt characteristic length is 1 mm. The whole patch size is 5 mm.

In addition to patch size effect, the dependence of velocity dispersion/attenuation on permeability is also explored. As the matrix permeability is reduced from 5 (Fig. 4) to 0.1 mD (Fig. 8), the velocity transition region and attenuation peaks of BIPS, BISQ, and TLP models shift toward low frequency (Fig. 8), whereas the Biot peak exhibits the opposite behavior. The physical meaning of this disparity has been attributed to the difference between Biot mechanism and squirt-flow mechanism.[71] It is obvious that the relative position and heights of BIPS, BISQ, and TLP peaks do not have apparent changes [Fig. 8(b)], which is consistent at different permeabilities.

The current analysis clearly shows a significant difference in the mechanism of Biot's theory, BISQ, and patchy-saturation. As one can see, TLP model and Biot's theory are the same in full saturation case since the LFIF effect of TLP model is canceled. At the same time, BIPS model reduces to BISQ model for the reason that only Biot viscous dissipation mechanism and squirt flow mechanism are retained for BIPS in full saturation case. It is obvious that in partial-saturation case, BIPS model persists in all the three (Biot-patchy-squirt) mechanisms, whereas BISQ model contains the Biot-squirt mechanism and TLP model contains the Biot-patchy mechanism. The effects of frequency, inner pocket fluid and permeability on peaks of different mechanism show distinctive characteristics, which strongly indicate that the Biot, patchy-saturation, and squirt flow mechanisms are the three fundamental physical processes that dominate the wave propagations in partially saturated double-porosity media. Any one of the three mechanisms does not provide adequate interpretations of the dispersion and attenuation in natural rocks.

In summary, Biot theory, patchy-saturation mechanism, and squirt flow mechanism are different from each other. In addition, they share common parts and couple with each other. These results, taken with the fact that matrix/fluid inhomogeneity is a quite typical phenomenon in natural rocks, indicate that the BIPS model can be regarded as a general model to predict dispersion and attenuation in reservoir rocks.

### C. Prediction of attenuation caused by squirt flow and LFIF flow in sandstone

Subramaniyan et al.[63] reported the evidence of squirt flow in Fontainebleau sandstone through laboratory observation of seismic attenuation. A Fontainebleau sandstone (porosity 8% and permeability 12 mD) was saturated with the glycerin–water mixture of varying viscosity (thus varying volume fraction of glycerin in the mixture). The Young's module attenuations were measured at the confining pressures of 5, 10, and 15 MPa. As the viscosity of the saturating fluid decreases (thus glycerin fraction decrease in the mixture), the attenuation magnitude drops and the peak frequency shifts to higher frequencies. The attenuation peak frequency is inversely proportional to fluid viscosity, which is a result of squirt flow.[73]

The squirt flow, patch saturation, and global fluid flow mechanism are clearly related, but in a way that is not always obvious. The relationship between attenuation and peak frequency in patchy-saturation model is similar to that of squirt flow, i.e., the peak frequency decreases as the viscosity increases. The key to understanding the BIPS model is to find out the reason why the attenuation frequency peak is inversely proportional to the viscosity.

There are two main causes of attenuation at low pressure: (1) LFIF at patch interface; (2) squirt flow between compliant and stiff pore. At the high pressure, all the compliant pores in the sandstone are assumed to be closed. Thus, the squirt flow effect will also cease. For glycerin–water mixture saturation, if attenuation magnitude drops and peak frequency shifts inversely proportional to fluid viscosity, the attenuation is no doubt caused by patchy-saturation.

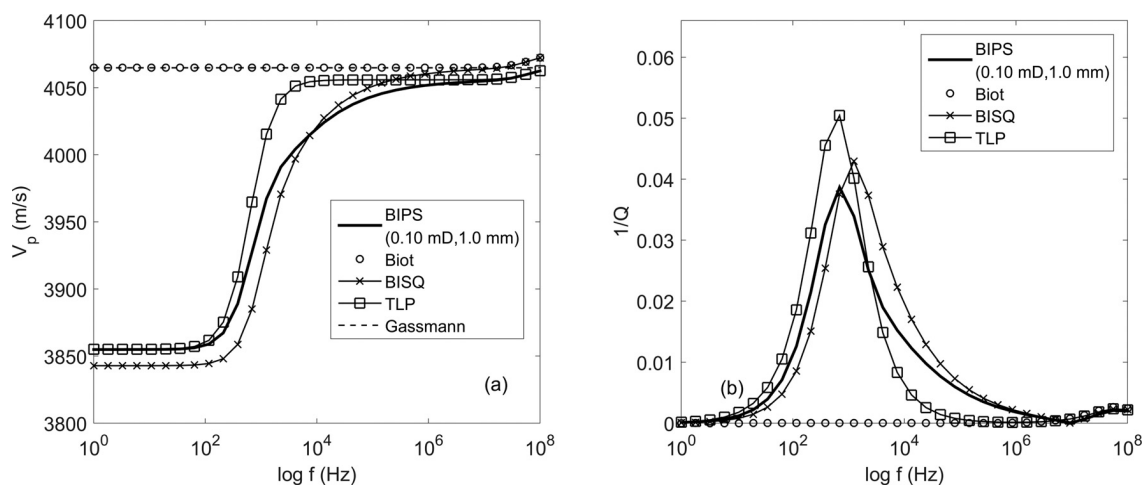

**FIG. 8.** Prediction of P-wave velocity dispersion (a) and attenuation (b) by BIPS, BISQ, TLP, Biot, and Gassmann methods. The volume fraction of inner fluid pocket (water) is 0.9. The permeability is 0.1 mD. The squirt characteristic length is 1 mm.

Subramaniyan et al. reported the attenuation at high pressure (15 MPa) in sandstone for glycerin full saturation.[73] In such case, there exists no patch interface and LFIF effect cannot be identified. Fortunately, the attenuations in partially saturated Berea sandstone at high pressure (15 MPa) have been reported recently,[74] which is caused by LFIF, rather than squirt flow. The Berea sandstone has a porosity of 18% and a permeability of 50 mD. Solid grain density is 2650 kg/m$^3$. Bulk modulus of solid grains is 36 GPa. Bulk and shear moduli of dry frame are 15.1 and 14 GPa (at confining pressure 25 MPa). Bulk modulus, density and viscosity of water are 2.2 GPa, 1000 kg/m$^3$, and 0.001 Pa s. Bulk modulus, density and viscosity of air are 0.1 MPa, 1 kg/m$^3$, and $2 \times 10^{-5}$ Pa s. The sandstone is fully saturated with water first. Then water saturation decreases to lower percentages. At confining pressure of 15 MPa, the extensional-mode attenuation magnitude was measured for 96%, 98%, 99%, and 100% water saturation (Fig. 9). Attenuation decreases and the peak shifts to higher frequency as water saturation decrease from nearly 100%–96%. Full water saturation indicates there is only one uniform fluid in pore space. As water saturation decreases, air patches appear and increase gradually. LFIF becomes more and more important as water-air interface expands.

Here attenuations have been predicted by BIPS model and compared with laboratory data (Fig. 9). Since the attenuations were measured at high confining pressure, compliant pore/cracks remain closed during the measurements. Thus, squirt flow effect is absence in the attenuation. The experiment data and predicted attenuations show clearly that the peak frequency is inversely proportional to water saturation (from 100%–96%), which is an evidence of LFIF. This phenomenon has also been observed in the numerical example in Figs. 4(b) and 5(b).

The results show that the BIPS model allow us to reproduce the attenuation-frequency relation observed in laboratory. Predicted attenuations exhibit the same dependence on water saturation as experimental data. It is important to recall that patch size must be considered as a function of saturation. Although it iss not easy to measure patch radius directly, it can be estimated by matching the calculated velocity with experiment data. Indeed, the patch radius has been estimated based on the four-term polynomial formula with coefficients.[56] Even though how the fluid patch evolves in porous media is still poorly understood, the application of quantitative function between patch radius and saturation in this study is expected to shed light on the remarkable role of partial-saturation in affecting wave propagations.

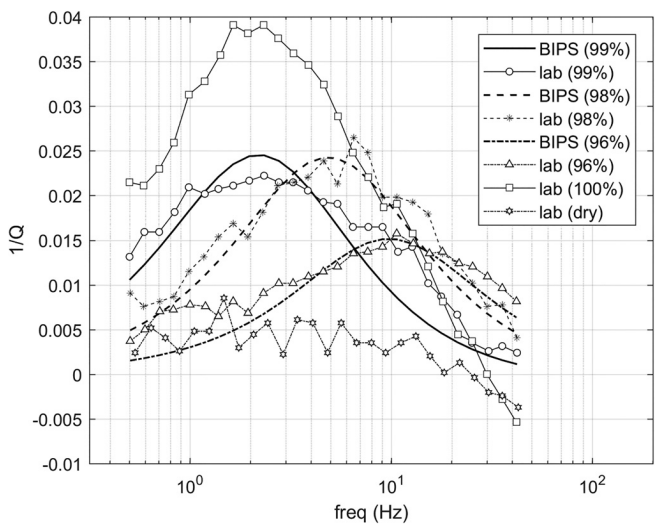

**FIG. 9.** Attenuation for confining pressures of 15 MPa in Berea sandstone (following Chapman et al., 2016). The water saturation ranging from dry conditions to around 100%. BIPS model prediction and the experiment data are shown in different lines.

## V. CONCLUSIONS

A Biot-patchy-squirt (BIPS) model is proposed, and wave equations are obtained for partially saturated double-porosity medium. Wave dispersion and attenuation caused by viscous dissipation, squirt flow, and patchy-saturation interface vibration are formulated simultaneously. The importance of current work is twofold. (1) BIPS model combines macroscopic global fluid flow, mesoscopic patchy heterogeneities, and microscopic squirt effect. The model makes it possible to study the coupling effects of the three energy dissipation mechanisms. (2) BIPS model incorporates both fluid and solid heterogeneities in porous media, which is commonly observed in natural reservoir rocks. The BIPS is a more general model that can be reduced to the Biot's theory, the BISQ model, and the triple-layer patchy (TLP) saturation model. Through numerical examples and analysis of effects of frequency, fluid type, patch, and permeability, it is confirmed that the Biot's viscous dissipation, patchy-saturation, and squirt flow are different wave dispersion/attenuation mechanism and are coupled to each other. The existence of squirt flow partially counterbalances the dissipative effect of LFIF at the patch interface. Therefore, in fractured and partially saturated rocks, no single mechanism is sufficient to explain wave dispersion/attenuation, and the combined effects play an important role in causing wave attenuation. The predicted attenuations are in good agreement with the experimental data, which confirms that the BIPS model can capture the relationship among the three mechanisms. This new method enables a deeper understanding of the wave phenomena in partially saturated inhomogeneous media.

## AUTHORS' CONTRIBUTIONS

All authors contributed equally to this work.

## ACKNOWLEDGMENTS

The research was supported by the National Natural Science Foundation of China (Grant Nos. 41874137 and 42074144) and National Key R&D Program of China (2018YFA0702501).

## APPENDIX A: CONSTITUTIVE RELATIONSHIP OF DOUBLE POROSITY POROUS MEDIUM

Assuming the porous medium is isotropic and pore fluids have no shear strains, the relationship among changes in total stress $d\boldsymbol{\sigma}$, solid stress $d\boldsymbol{\sigma}_s$, and fluid pressure $dp$ is governed by[69,75,76]

$$d\boldsymbol{\sigma}_1 = d\boldsymbol{\sigma}_{s1} - \alpha dp_1 \mathbf{I}, \quad (A1)$$

$$d\boldsymbol{\sigma}_2 = d\boldsymbol{\sigma}_{s2} - \alpha dp_2 \mathbf{I}, \quad (A2)$$

where **I** is the unit tensor and $\alpha = \frac{2(1+\nu)G}{3(1-2\nu)H}$. The constants $G, \nu$ are the shear modulus and Poisson's ratio. The coefficient $1/H$ is a measure of the solid matrix compressibility for a change in water pressure.[69,70] Here, subscripts 1 and 2 refer to phases with different porosities.

In the case of liner elasticity, the stress-strain relationships of solid phases are

$$d\boldsymbol{\sigma}_{s1} = \mathbf{M}_1 d\mathbf{e}_1, \quad (A3)$$

$$d\boldsymbol{\sigma}_{s2} = \mathbf{M}_2 d\mathbf{e}_2, \quad (A4)$$

where **e** is the solid strain and **M** is the stiffness matrix. The total strain in double porosity matrix due to strains in each phase is

$$\int_{\Omega_1+\Omega_2} d\mathbf{e} dv = \int_{\Omega_1} d\mathbf{e}_1 dv + \int_{\Omega_2} d\mathbf{e}_2 dv$$
$$= \int_{\Omega_1} \mathbf{M}_1^{-1}(d\boldsymbol{\sigma}_1 + \alpha dp_1 \mathbf{I}) dv + \int_{\Omega_2} \mathbf{M}_2^{-1}(d\boldsymbol{\sigma}_2 + \alpha dp_2 \mathbf{I}) dv, \quad (A5)$$

where $\Omega_m$ is the space of phase $m$. Considering the equilibrium condition of local stress, the stress changes in nearby phases satisfy $d\boldsymbol{\sigma}_1 = d\boldsymbol{\sigma}_2 = d\bar{\boldsymbol{\sigma}}$. If the local strain in phase $m$ is homogeneous, the integrals are to be replaced by volume fraction $v_1, v_2$. The total strain is

$$\mathbf{e} = v_1 \mathbf{e}_1 + v_2 \mathbf{e}_2 = \left(v_1 \mathbf{M}_1^{-1} + v_2 \mathbf{M}_2^{-1}\right) \bar{\boldsymbol{\sigma}} + v_1 \alpha \mathbf{M}_1^{-1} p_1 \mathbf{I} + v_2 \alpha \mathbf{M}_2^{-1} p_2 \mathbf{I} \quad (A6)$$

or

$$\bar{\boldsymbol{\sigma}} = \left(v_1 \mathbf{M}_1^{-1} + v_2 \mathbf{M}_2^{-1}\right)^{-1} \left(\mathbf{e} - v_1 \alpha \mathbf{M}_1^{-1} p_1 \mathbf{I} - v_2 \alpha \mathbf{M}_2^{-1} p_2 \mathbf{I}\right). \quad (A7)$$

For uniaxial deformation in the $x$-direction ($e_y = e_z = 0$), total stress satisfies

$$\bar{\sigma}_x = \bar{M} \frac{du}{dx} - \alpha_1' p_1 - \alpha_2' p_2, \quad (A8)$$

where $\bar{M} = \frac{M_1 M_2}{v_1 M_2 + v_2 M_1}$, $\alpha_1' = \frac{v_1 \bar{M}}{M_1} \alpha$, $\alpha_2' = \frac{v_2 \bar{M}}{M_2} \alpha$, and $M_m = 2G_m \frac{1-\nu_m}{1-2\nu_m}$, $m=1,2$.

The relationship among pore fluid pressure $p$, solid strain $\mathbf{e} = (e_x, e_y, e_z)^T$, and fluid content variation $\boldsymbol{\xi} = (\xi_1, \xi_2)^T$ is governed by[14,17]

$$s_1 = R_1 \xi_1 + \sigma_1', \quad (A9)$$

$$s_2 = R_2 \xi_2 + \sigma_2', \quad (A10)$$

where the stress on fluid is $s_m = \phi_m p_m$ and $\sigma_m' = Q_m I_1$, $m=1,2$. $I_1 = e_x + e_y + e_z$, and $R_m, Q_m$ are the elastic constant as defined in Biot's work.[17] $\sigma_m'$ is a measure of the change in pore fluid pressure for solid strain. At long-term equilibrium, the change in fluid pressure relax to a stable value $\sigma'$. Without losing generality, let $\sigma_1' = \sigma' = \sigma_2'/\beta^*$ and $\beta^* = Q_2/Q_1$.

The total fluid content variation in each of the phase is

$$\xi = v_1 \xi_1 + v_2 \xi_2 = -\left(\frac{v_1}{R_1} + \frac{v_2}{R_2}\beta\right)\sigma' + \frac{v_1}{R_1}s_1 + \frac{v_2}{R_2}s_2 \quad (A11)$$

or

$$R'\xi + \sigma' = \frac{R'}{R_1}v_1 s_1 + \frac{R'}{R_2}v_2 s_2, \quad (A12)$$

where $R' = \frac{R_1 R_2}{v_1 R_2 + v_2 R_1 \beta^*}$. The stress on the fluid caused by fluid content variation in each phase is defined as $s' = R'\xi + \sigma'$. The total stress of porous medium is separated into solid part $\sigma_x$ and fluid part $s'$,

$$\bar{\sigma}_x = \sigma_x - s' = \sigma_x - \frac{R'}{R_1}v_1 \sum_{m=1}^{2}\phi_m p_1 - \frac{R'}{R_2}v_2 \sum_{m=1}^{2}\phi_m p_2. \quad (A13)$$

Here $\bar{\sigma}_x$ is the total stress in the $x$ direction. Combining (A8) and (A13), one can easily have

$$\sigma_x = \bar{M}\frac{du}{dx} - \gamma_1 p_1 - \gamma_2 p_2, \quad (A14)$$

where $\gamma_1 = \frac{\bar{M}}{M_1}v_1 \alpha - \frac{R'}{R_1}v_1 \sum_{m=1}^{2}\phi_{m1}$, $\gamma_2 = \frac{\bar{M}}{M_2}v_2 \alpha - \frac{R'}{R_2}v_2 \sum_{m=1}^{2}\phi_m$, $\bar{M} = \frac{M_1 M_2}{v_1 M_2 + v_2 M_1}$.

Considering following relations:

$$\frac{\bar{M} v_1}{M_1} = \frac{v_1 M_2}{v_1 M_2 + v_2 M_1} = \frac{v_1 \phi_{10}}{v_1 \phi_{10} + v_2 \phi_{20}\frac{\phi_{10}M_1}{\phi_{20}M_2}} = \frac{\phi_1}{\phi_1 + \phi_2 \frac{\phi_{10}M_1}{\phi_{20}M_2}}, \quad (A15)$$

$$\frac{R' v_1}{R_1} = \frac{v_1 R_2}{v_1 R_2 + v_2 R_1 \beta^*} = \frac{\phi_1}{\phi_1 + \phi_2 \frac{\phi_{10}R_1\beta^*}{\phi_{20}R_2}}, \quad (A16)$$

the coefficients $\gamma_1, \gamma_2$ can be rewritten as

$$\gamma_1 = \left(\alpha - \sum_{m=1}^{2}\phi_m\right)\frac{\phi_1}{\phi_1 + \phi_2 \beta}, \quad (A17)$$

$$\gamma_2 = \left(\alpha - \sum_{m=1}^{2}\phi_m\right)\frac{\phi_2}{\phi_2 + \phi_1/\beta}, \quad (A18)$$

where

$$\beta = \frac{M_1 \phi_{10}}{M_2 \phi_{20}} = \frac{R_1 \phi_{10}}{R_2 \phi_{20}}\beta^*. \quad (A19)$$

The parameter $\alpha$ is defined as $\alpha = \frac{2(1+\nu)G}{3(1-2\nu)Q} = \frac{K}{Q}$[61,69] where $\frac{1}{Q} = \frac{1}{K} - \frac{1}{K_s}$. Here, $v_1, v_2$ are the volume fraction of the phase 1 (inner pocket) and phase 2 (surrounding patch), respectively. Here, $\nu$ is Poisson's ratio, $G$ is shear modulus, and $K$ is bulk modulus.

The conservation equations in partially saturated double-porosity medium are given as

$$\frac{\partial(\rho_{f_1}\phi_1)}{\partial t} + \frac{\partial\left[\rho_{f_1}\left(q_x^{(1)} - \phi_1 \dot{u}_x\right)\right]}{\partial x} + \frac{\partial\left(\rho_{f_1}q_r^{(1)}\right)}{\partial r} + \frac{\rho_{f_1}q_r^{(1)}}{r} = 0,$$

$$\frac{\partial(\rho_{f_2}\phi_2)}{\partial t} + \frac{\partial\left[\rho_{f_2}\left(q_x^{(2)} - \phi_2 \dot{u}_x\right)\right]}{\partial x} + \frac{\partial\left(\rho_{f_2}q_r^{(2)}\right)}{\partial r} + \frac{\rho_{f_2}q_r^{(2)}}{r} = 0,$$

$$(A20)$$

where $q_x^{(1)} = \phi_1(\dot{U}_x^{(1)} - \phi_2 \dot{Z}_x)$ and $q_x^{(2)} = \phi_2(\dot{U}_x^{(2)} + \phi_1 \dot{Z}_x)$ are the volumetric flux rate in the $x$-directions within the fluid patches. $Z_x$

is the $x$-direction component of the local fluid flow between different fluid patches. Here, $q_r^{(1)} = \phi_1 \dot{U}_r^{(1)}$ and $q_r^{(2)} = \varphi_2 \dot{U}_r^{(2)}$ are in the $r$-directions. Linearization gives

$$\frac{\phi_1}{\rho_{f_1}} \frac{\partial \rho_{f_1}}{\partial t} + \frac{\partial \phi_1}{\partial t} + \phi_1 \frac{\partial^2 (U_x^{(1)} - \phi_2 Z_x - u_x)}{\partial t \partial x}$$
$$+ \phi_1 \left( \frac{\partial^2 (U_r^{(1)})}{\partial t \partial r} + \frac{1}{r} \frac{\partial (U_r^{(1)})}{\partial t} \right) = 0,$$
$$\frac{\phi_2}{\rho_{f_2}} \frac{\partial \rho_{f_2}}{\partial t} + \frac{\partial \phi_2}{\partial t} + \phi_2 \frac{\partial^2 (U_x^{(2)} + \phi_1 Z_x - u_x)}{\partial t \partial x}$$
$$+ \phi_2 \left( \frac{\partial^2 (U_r^{(2)})}{\partial t \partial r} + \frac{1}{r} \frac{\partial (U_r^{(2)})}{\partial t} \right) = 0. \quad (A21)$$

The variation in fluid content, i.e., the porosity change, is $d\phi_m = \alpha \frac{\partial u}{\partial x} + \frac{p_m}{H'_m}$, ($x = 1, 2$). Here, $H'_1 = H_1 + H_1 \phi_2 \beta / \phi_1$, $H'_2 = H_2 + H_2 \phi_1 / (\beta \phi_2)$, $1/H_m = (1 - \sum_{m=1}^{2} \phi_m - K/K_s)/K_s$. The porosity differential is

$$\frac{d\phi_m}{dt} = \alpha \frac{\partial^2 u_x}{\partial x \partial t} + \frac{1}{H'_m} \frac{\partial p_m}{\partial t}. \quad (A22)$$

Here, $d\phi_m (m = 1, 2)$ is the porosity change, i.e., the variation in fluid content. The fluid density compressibility is represented as $d\rho_f = \frac{dp}{c_0^2}$, where $c_0$ is acoustic velocity of fluid. By substituting $\frac{d\phi}{dt}$ and $d\rho_f$ into Eq. (A21), the equations governing fluid pressure are

$$\frac{dp_1}{dt} = -F_1 \left( \frac{\partial^2 (U_x^{(1)} - \phi_2 Z_x)}{\partial t \partial x} + \frac{\partial^2 (U_r^{(1)})}{\partial t \partial r} + \frac{1}{r} \frac{\partial (U_r^{(1)})}{\partial t} + \frac{\gamma_1}{\phi_1} \frac{\partial^2 u_x}{\partial t \partial x} \right),$$
$$\frac{dp_2}{dt} = -F_2 \left( \frac{\partial^2 (U_x^{(2)} + \phi_1 Z_x)}{\partial t \partial x} + \frac{\partial^2 (U_r^{(2)})}{\partial t \partial r} + \frac{1}{r} \frac{\partial (U_r^{(2)})}{\partial t} + \frac{\gamma_2}{\phi_2} \frac{\partial^2 u_x}{\partial t \partial x} \right),$$
$$(A23)$$

where $F_1 = \frac{K_{f_1}(\phi_1 + \phi_2 \beta) H_1}{K_{f_1} + (\phi_1 + \phi_2 \beta) H_1}$, $F_2 = \frac{K_{f_2}(\phi_2 + \phi_1 / \beta) H_2}{K_{f_2} + (\phi_2 + \phi_1 / \beta) H_2}$.

There are eight unknowns, $U_x^{(1)}, U_x^{(2)}, U_r^{(1)}, U_r^{(2)}, u_x, Z_x, p_1, p_2$, and eight differential equations. In the case of plane wave propagation, the displacements and pressure are expressed by wave kernel $e^{i(\omega t - \mathbf{k} \cdot \mathbf{x})}$,

$$u_x(x, t) = C_1 e^{i(\omega t - \mathbf{k} \cdot \mathbf{x})}, \quad U_x^{(1)}(x, t) = C_2^{(1)} e^{i(\omega t - \mathbf{k} \cdot \mathbf{x})},$$
$$U_x^{(2)}(x, t) = C_2^{(2)} e^{i(\omega t - \mathbf{k} \cdot \mathbf{x})}, \quad Z_x(x, t) = C_3 e^{i(\omega t - \mathbf{k} \cdot \mathbf{x})},$$
$$U_r^{(1)}(x, r, t) = U_{r0}^{(1)}(r) e^{i(\omega t - \mathbf{k} \cdot \mathbf{x})}, \quad U_r^{(2)}(x, r, t) = U_{r0}^{(2)}(r) e^{i(\omega t - \mathbf{k} \cdot \mathbf{x})},$$
$$p_1(x, r, t) = p_{01}(r) e^{i(\omega t - \mathbf{k} \cdot \mathbf{x})}, \quad p_2(x, r, t) = p_{02}(r) e^{i(\omega t - \mathbf{k} \cdot \mathbf{x})}.$$
$$(A24)$$

Then fluid pressures satisfy

$$\frac{\partial p_{01}}{\partial r} = U_{r0}^{(1)} \rho_{f_1} \omega^2 \left( \frac{1 + 1/\phi_{10}}{2} - i \frac{\omega_c^{(1)}}{\omega} \right),$$
$$\frac{\partial p_{02}}{\partial r} = U_{r0}^{(2)} \rho_{f_2} \omega^2 \left( \frac{1 + 1/\phi_{20}}{2} - i \frac{\omega_c^{(2)}}{\omega} \right), \quad (A25)$$

where $\omega_c^{(m)} = \frac{\mu_m \phi_{m0}}{\rho_{fm} \kappa_m}$, $m = 1, 2$.

The ordinary differential equations governing $p_{01}$ and $p_{02}$ are

$$\frac{\partial^2 p_{01}}{\partial r^2} + \frac{1}{r} \frac{\partial p_{01}}{\partial r} + \frac{\rho_{f_1} \omega^2}{F_1} \left( \frac{1 + 1/\phi_{10}}{2} - i \frac{\omega_c^{(1)}}{\omega} \right) p_{01}$$
$$- i k \rho_{f_1} \omega^2 \left( C_2^{(1)} - \phi_2 C_3 + \frac{\gamma_1}{\phi_1} C_1 \right) \left( \frac{1 + 1/\phi_{10}}{2} - i \frac{\omega_c^{(1)}}{\omega} \right) = 0,$$
$$\frac{\partial^2 p_{02}}{\partial r^2} + \frac{1}{r} \frac{\partial p_{02}}{\partial r} + \frac{\rho_{f_2} \omega^2}{F_2} \left( \frac{1 + 1/\phi_{20}}{2} - i \frac{\omega_c^{(2)}}{\omega} \right) p_{02}$$
$$- i k \rho_{f_2} \omega^2 \left( C_2^{(2)} + \phi_1 C_3 + \frac{\gamma_2}{\phi_2} C_1 \right) \left( \frac{1 + 1/\phi_{20}}{2} - i \frac{\omega_c^{(2)}}{\omega} \right) = 0.$$
$$(A26)$$

## APPENDIX B: PLANE WAVE ANALYSIS EQUATIONS

The coefficients in P-wave dispersion equation are

$$S_1 = \frac{\phi_2 \phi_2 \phi_1 (\chi_1 \omega^2 - i \beta_1 \omega)}{3} + \phi_1 \phi_2 (\tilde{F}_1 \phi_2 + \tilde{F}_2 \phi_1), \quad (B1)$$

$$S_{10} = \frac{\phi_2 \phi_{20} \phi_{10} (\chi_1 \omega^2 - i \beta_1 \omega)}{3} + \phi_{10} \phi_{20} (\tilde{F}_1 \phi_2 + \tilde{F}_2 \phi_1), \quad (B2)$$

$$S_2 = \tilde{F}_2 \gamma_{20} \phi_1 \nu_2 - \tilde{F}_1 \gamma_{10} \phi_2 \nu_1, \quad (B3)$$

$$S_{20} = \tilde{F}_2 \gamma_{20} \phi_{10} - \tilde{F}_1 \gamma_{10} \phi_{20}, \quad (B4)$$

$$a_{11} = \bar{M} - \frac{S_2 S_{20}}{S_{10}} + \frac{\tilde{F}_1 \gamma_1 \gamma_{10}}{\phi_{10}} + \frac{\tilde{F}_2 \gamma_2 \gamma_{20}}{\phi_{20}}, \quad (B5)$$

$$a_{12} = a_{21} = \tilde{F}_1 \gamma_1 + \frac{\tilde{F}_1 \phi_{10} \phi_{20} S_2}{S_{10}}, \quad (B6)$$

$$a_{13} = a_{31} = \tilde{F}_2 \gamma_2 + \frac{\tilde{F}_2 \phi_{10} \phi_{20} S_2}{S_{10}}, \quad (B7)$$

$$a_{22} = \tilde{F}_1 \phi_1 \left( 1 - \frac{\tilde{F}_1 \phi_{10} \phi_{20} \phi_2}{S_{10}} \right), \quad (B8)$$

$$a_{23} = a_{32} = \frac{\tilde{F}_1 \tilde{F}_2 \phi_{10} \phi_{20} \phi_1 \phi_2}{S_{10}}, \quad (B9)$$

$$a_{33} = \tilde{F}_2 \phi_2 \left( 1 - \frac{\tilde{F}_2 \phi_{10} \phi_{20} \phi_1}{S_{10}} \right), \quad (B10)$$

$$b_{11} = -\rho_{00} + \frac{i(b_1 + b_2)}{\omega}, \quad (B11)$$

$$b_{12} = b_{21} = -\rho_{01} - \frac{ib_1}{\omega}, \quad (B12)$$

$$b_{13} = b_{31} = -\rho_{02} - \frac{ib_2}{\omega}, \quad (B13)$$

$$b_{22} = -\rho_{11} + \frac{ib_1}{\omega}, \quad (B14)$$

$$b_{23} = b_{32} = 0, \quad (B15)$$

$$b_{33} = -\rho_{22} + \frac{ib_2}{\omega}, \quad (B16)$$

where $\tilde{F}_m = F_m \left( 1 - \frac{2 J_1(\lambda_m R)}{\lambda_m R J_0(\lambda_m R)} \right)$, $F_1 = \frac{K_{f_1}(\phi_1 + \phi_2 \beta) H_1}{K_{f_1} + (\phi_1 + \phi_2 \beta) H_1}$, $F_2 = \frac{K_{f_2}(\phi_2 + \phi_1/\beta) H_2}{K_{f_2} + (\phi_2 + \phi_1/\beta) H_2}$. The relations $\phi_1 = \phi_{10} \nu_1$, $\phi_2 = \phi_{20} \nu_2$, $\gamma_1 = \gamma_{10} \nu_1$, $\gamma_2 = \gamma_{20} \nu_2$,

$\gamma_{10} = (\alpha - \sum_{m=1}^{2} \phi_m)\frac{\phi_{10}}{\phi_1+\phi_2\beta}$, and $\gamma_{20} = (\alpha - \sum_{m=1}^{2} \phi_m)\frac{\phi_{20}}{\phi_2+\phi_1/\beta}$ are used in deriving the coefficients.

The coefficients $A, B, C, D$ are

$$A = a_{11}a_{22}a_{33} - a_{11}a_{23}a_{32} - a_{12}a_{21}a_{33} + a_{12}a_{23}a_{31} + a_{13}a_{21}a_{32} - a_{13}a_{22}a_{31}, \tag{B17}$$

$$\begin{aligned}B =\ & a_{11}a_{22}b_{33} - a_{11}a_{23}b_{32} - a_{11}a_{32}b_{23} + a_{11}a_{33}b_{22} \\ & - a_{12}a_{21}b_{33} + a_{12}a_{23}b_{31} + a_{12}a_{31}b_{23} - a_{12}a_{33}b_{21} \\ & + a_{13}a_{21}b_{32} - a_{13}a_{22}b_{31} - a_{13}a_{31}b_{22} + a_{13}a_{32}b_{21} \\ & + a_{21}a_{32}b_{13} - a_{21}a_{33}b_{12} - a_{22}a_{31}b_{13} + a_{22}a_{33}b_{11} \\ & + a_{23}a_{31}b_{12} - a_{23}a_{32}b_{11}, \end{aligned} \tag{B18}$$

$$\begin{aligned}C =\ & a_{11}b_{22}b_{33} - a_{11}b_{23}b_{32} - a_{12}b_{21}b_{33} + a_{12}b_{23}b_{31} \\ & + a_{13}b_{21}b_{32} - a_{13}b_{22}b_{31} - a_{21}b_{12}b_{33} + a_{21}b_{13}b_{32} \\ & + a_{22}b_{11}b_{33} - a_{22}b_{13}b_{31} - a_{23}b_{11}b_{32} + a_{23}b_{12}b_{31} \\ & + a_{31}b_{12}b_{23} - a_{31}b_{13}b_{22} - a_{32}b_{11}b_{23} + a_{32}b_{13}b_{21} \\ & + a_{33}b_{11}b_{22} - a_{33}b_{12}b_{21}, \end{aligned} \tag{B19}$$

$$\begin{aligned}D =\ & b_{11}b_{22}b_{33} - b_{11}b_{23}b_{32} - b_{12}b_{21}b_{33} + b_{12}b_{23}b_{31} \\ & + b_{13}b_{21}b_{32} - b_{13}b_{22}b_{31}. \end{aligned} \tag{B20}$$

## DATA AVAILABILITY

The data that support the findings of this study are available from the corresponding author upon reasonable request.